\documentclass{an_art}
\usepackage{graphicx}

\begin{document}

\begin{titlepage}
\setcounter{page}{1}
\headnote{Astron.~Nachr.~320 (1999) 1, 21-25}
\makeheadline
\title {Extragalactic Cepheid database}

\author{
{\sc P.~Lanoix,
R.~Garnier, 
G.~Paturel, 
C.~Petit, 
J.~Rousseau {\rm and}
H.~Di Nella-Courtois}, Lyon, France \\
\medskip
{\small Centre de Recherche Astronomique de Lyon} \\
\bigskip
}

\date{Received 1998 October 23; accepted 1999 February 3}
\maketitle

\summary
We present in this paper an exhaustive compilation of all published data of
extragalactic Cepheids. We have checked every light curve in order to
characterize the different types of Cepheid and detect potential overtone
pulsators, or to estimate the quality of the data. This
compilation of about 3000 photometric measurements will constitute a very
useful tool for astronomers involved for instance in the extragalactic distance scale.
END

\keyw
Database - Cepheids: extragalactic
END
\end{titlepage}

\section{Introduction}
It is well known that Cepheid pulsating stars constitute, through the
Period-Luminosity (PL) relation, the one primary calibrator
for extragalactic distance scale. Photometric data of extragalactic
Cepheids are then the raw material 
of its calibration, and, thus, it is of greater importance to have as much
data as possible.

With the aim of computing new distances based on our direct calibration
of the PL relation (Lanoix et al. 1999a) while taking into account the
PL relation incompleteness bias (Lanoix et al. 1999b) first, and subsequently 
calibrating secondary distance 
calibrators such as Tully-Fisher relation, luminosity peak brightness of SNe
Ia (Lanoix 1998) or Faber-Jackson relation for globular clusters (Di
Nella-Courtois et al. 1999), 
we have looked for photometric data of extragalactic Cepheids. We found that
Madore published in 1985 a compilation of all data available at this time.
However, this compilation actually put data from 10 galaxies together, and
to our knowledge, no updated version of this paper exists,
whereas data from more than 30 galaxies is now available.

We have therefore updated this compilation for our own work while putting 
quality flags to every light curve,
and have now decided to put this useful tool at the
astronomical community's disposal.

\section{Data} 
We collect 3031 photometric measurements of 1061 Cepheids located in 
33 galaxies (without including SMC and LMC). Table \ref{galaxy} gives
the complete list and the main characteristic of those galaxies according to 
the LEDA database (http://www-obs.univ-lyon1.fr/leda/home$\_$leda.html).
Our bibliography
is as exhaustive as possible, and is complete until November 1998, while
new publications are still arriving. 

\begin{table}[ht]
\caption{List of the 33 galaxies of the database}
\label{galaxy}
\begin{center}
{\small
\begin{tabular}{llrrlcrc}\hline
Name & PGC/LEDA number & RA 2000 & DEC 2000 & Morph. type & Lum. class code & Total B-magnitude & HST \\
\hline
DDO 155  & PGC 44491 &   12.97775& 14.21618 &   Irr    &  9.000  & 14.715 & n  \\
DDO 216  & PGC 71538 &   23.47614& 14.74660 &   Irr    &  9.000  & 12.789 & n  \\
DDO 50   & PGC 23324 &    8.31831& 70.71419 &   Irr    &  8.279  & 11.092 & n  \\
DDO 69   & PGC 28868 &    9.98995& 30.74495 &   Irr    &  9.000  & 12.956 & n  \\
IC 10    & PGC 01305 &     .34016& 59.29171 &   Irr    &  9.000  & 12.197 & n  \\
IC 1613  & PGC 03844 &    1.08172&  2.13330 &   Irr    &  9.230  &  9.933 & n  \\
IC 4182  & PGC 45314 &   13.09704& 37.60582 &   Sm     &  8.338  & 12.409 & y  \\
NGC 1365 & PGC 13179 &    3.56016&-36.13807 &   SBb    &  1.371  & 10.350 & y  \\
NGC 2090 & PGC 17819 &    5.78398&-34.25145 &   Sc     &  3.931  & 11.767 & y  \\
NGC 224  & PGC 02557 &     .71232& 41.26897 &   Sb     &  2.000  &  4.170 & n  \\
NGC 2366 & PGC 21102 &    7.48175& 69.21442 &   Irr    &  8.722  & 11.430 & n  \\
NGC 2403 & PGC 21396 &    7.61513& 65.59957 &   SBc    &  5.000  &  8.824 & n  \\
NGC 2541 & PGC 23110 &    8.24451& 49.06227 &   SBc    &  6.696  & 12.043 & y  \\
NGC 300  & PGC 03238 &     .91493&-37.68250 &   Scd    &  5.969  &  8.785 & n  \\
NGC 3031 & PGC 28630 &    9.92597& 69.06665 &   Sb     &  2.000  &  7.687 & y  \\
NGC 3109 & PGC 29128 &   10.05185&-26.15890 &   SBm    &  7.924  & 10.347 & n  \\
NGC 3351 & PGC 32007 &   10.73278& 11.70408 &   SBb    &  3.000  & 10.382 & y  \\
NGC 3368 & PGC 32192 &   10.77922& 11.82098 &   SBab   &  3.000  &  9.916 & y  \\
NGC 3621 & PGC 34554 &   11.30466&-32.81352 &   SBcd   &  5.849  & 10.077 & y  \\
NGC 4321 & PGC 40153 &   12.38200& 15.82293 &   SBbc   &  1.000  &  9.992 & y  \\
NGC 4414 & PGC 40692 &   12.44097& 31.22479 &   Sc     &  3.576  & 10.923 & y  \\
NGC 4496A& PGC 41471 &   12.52771&  3.93911 &   SBd    &  5.600  & 12.116 & y  \\
NGC 4536 & PGC 41823 &   12.57414&  2.18848 &   SBbc   &  2.824  & 11.012 & y  \\
NGC 4725 & PGC 43451 &   12.84079& 25.50030 &   SBab   &  1.689  &  9.955 & y  \\
NGC 5253 & PGC 48334 &   13.66551&-31.64477 &   S?     &   /     & 10.765 & y  \\
NGC 5457 & PGC 50063 &   14.05356& 54.35075 &   SBc    &  1.000  &  8.197 & y  \\
NGC 598  & PGC 05818 &    1.56414& 30.66017 &   Sc     &  4.000  &  6.193 & n  \\
NGC 6822 & PGC 63616 &   19.74940&-14.80306 &   Irr    &  8.493  &  9.322 & n  \\
NGC 7331 & PGC 69327 &   22.61809& 34.41949 &   Sbc    &  2.000  & 10.165 & y  \\
NGC 925  & PGC 09332 &    2.45467& 33.57817 &   SBcd   &  4.000  & 10.583 & y  \\
SEXTANS A& PGC 29653 &   10.18369& -4.71346 &   Irr    &  9.704  & 11.745 & n  \\
SEXTANS B& PGC 28913 &    9.99996&  5.33256 &   Irr    &  8.117  & 11.834 & n  \\
WLM      & PGC 00143 &     .03246&-15.45032 &   Irr    &  8.258  & 11.113 & n  \\
\hline
\end{tabular}}
\end{center}
\end{table}

One can note that some old photographic data has been rejected.
For instance, 
concerning NGC 224 (M31), we exclude data from Gaposchkin (1962) and Baade
\& Swope (1963, 1964) from our base. Moreover, we also exclude measurements
of that galaxy taken in very crowded fields, such as part 
of Welch et al. (1986) ones.  However finding charts or light curves
may be found in these papers.

Actually, the compilation can be divided into two subclasses: 
ground based data and {\it Hubble Space Telescope} (HST) data.
The first class appears very heterogeneous in many aspects (methods, 
limit magnitudes, bandpasses, time coverage, quality) since it's made of
observations from many different telescopes.
Moreover almost all those Cepheids were observed during a
single observation campaign too, so
that they cannot be compared to any other campaign.
On the other hand, the spatial
observations of the HST are highly homogeneous and
are composed of Cepheids from 17 galaxies at present.
However, concerning papers from the HST Key Project group (see
Freedman et al. 1994), we
have chosen to keep only data from ALLFRAME software package, while 
two sets of photometry may be available.
We have checked every
light curve (except for NGC 3368, NGC 4725 and NGC 224) in order
to allocate a flag to them,
and then to characterize the type of the corresponding Cepheid, or to give an
idea of the reliability of its photometry. Table \ref{quality}
describes the signification  of the different flags we use.

The ground-based measurements are divided up among 11 different bandpasses,
from B (440 nm) to K (2200 nm), whereas HST observed only in V and I bands.
Table \ref{band} gives the relation between notations and wavelengths.

\begin{table}[ht]
\caption{Bandpasses}
\label{band}
\begin{center}
{\small
\begin{tabular}{llll}\hline
Band & $\lambda_{eff}$ & Band & $\lambda_{eff}$ \\ 
     & (nm)            &      &  (nm)           \\
\hline \\
 B     &      440  & IV    &     1050 \\
 V     &      550  & J     &     1250  \\
 r     &      650  & H     &     1650 \\
 R     &      700  & K     &     2200 \\
 i     &      820  & g     &     500  \\
 I     &      900 \\
\hline
\end{tabular}}
\end{center}
\end{table}

\begin{table}[ht]
\caption{Description of light curves flags}
\label{quality}
\begin{center}
{\small
\begin{tabular}{ll}\hline
Flag & Light curve description \\
\hline \\
     N   &      Normal \\
     S   &      Symetrical (but high amplitude)\\
     B   &      Bumpy \\
     B+  &      Scattered or very bumpy \\
     O   &      Overtone\\
     O-  &      Low amplitude (but asymmetrical or with high period)\\
     P   &      Peculiar\\
     /   &      No curve\\
\hline
\end{tabular}}
\end{center}
\end{table}

Our base contains several types of magnitude corresponding to the calculation
method that is described in table \ref{type}. In some cases the sign ``:'' may
follow magnitudes or periods : it means that these magnitudes 
are doubtful and that these periods are just estimates or lower bounds of
the real values (according
to the original authors).

\begin{table}[ht]
\caption{Magnitudes description}
\label{type}
\begin{center}
{\small
\begin{tabular}{cl}\hline
Flag & Description \\
\hline \\
     mea    & Intensity averaged (based on curve area) \\
     max    & Maximum \\
     min    & Minimum\\
     ave    & Average of minimum and maximum\\
     /      & Single measurement \\
\hline
\end{tabular}}
\end{center}
\end{table}

The reference codes and the corresponding authors are given 
in table \ref{ref}. They can allow
the interested reader to look at the finding charts as well as the 
light curves.

Finally, the Cepheid names are those chosen by the authors, except in the case
of several stars in NGC 224 (Freedman \& Madore 1990, field IV)
and NGC 3368 (Tanvir 1995), where we
obtain data right from their figures and called them FRE -- -- and TAN -- --,
respectively.

\section{Table structure}
Data is presented as a table of more than 5000 lines according to the
structure described below for each Cepheid : \begin{itemize}
\item First line: Host galaxy, Cepheid name, number of following data lines for
that Cepheid
 \item Second line: pointer = 1, logarithm of the period in days, light
curve flag 
\item Next lines: pointer = 2, magnitude, type, band, reference
\end{itemize}

This database is available as an ASCII table on request by sending an
e-mail to Lanoix@obs.univ-lyon1.fr .

\newpage

\begin{table}[ht]
\caption{References}
\label{ref}
\begin{center}
{\small
\begin{minipage}[t]{70mm}
\begin{tabular}{ll}\hline
Reference code & Corresponding authors \\
\hline \\
Ala83 & Mc Alary et al. 1983 \\
Ala84 & Mc Alary et al. 1984 \\
Alv95 & Alves \& Cook 1995 \\ 
Car90 & Carlson \& Sandage 1990 \\      
Cap92 & Capaccioli et al. 1992 \\
Chr87 & Christian \& Schommer 1987 \\       
Coo86 & Cook \& Aaronson 1986 \\
Fer96 & Ferrarese et al. 1996 \\
Fer98 & Ferrarese et al. 1998 \\
Fr88a & Freedman 1988 \\                        
Fr88b & Freedman \& Madore 1988 \\
Fre90 & Freedman \& Madore 1990 \\
Fre91 & Freedman et al. 1991 \\
Fre92 & Freedman et al. 1992 \\
Fre94 & Freedman et al. 1994 \\             
Gal96 & Gallart et al. 1996 \\
Gib98 & Gibson et al. 1998 \\
Gra97 & Graham et al. 1997 \\ 
Hoe90 & Hoessel et al. 1990 \\
Hoe94 & Hoessel et al. 1994 \\
Hoe98 & Hoessel et al. 1998 \\
Hug98 & Hughes et al. 1998 \\
Kay67 & Kayser 1967 \\            
Kel96 & Kelson et al. 1996 \\
Kin87 & Kinman et al. 1987 \\
Mad85 & Madore et al. 1985 \\
\hline
\end{tabular}
\end{minipage}
\      
\begin{minipage}[t]{70mm}
\begin{tabular}{ll}\hline
Reference code & Corresponding authors  \\
\hline \\
Mad87 & Madore et al. 1987 \\
McA84 & Mc Alary \& Madore 1984\\
Mou87 & Mould 1987               \\             
Mus98 & Musella et al. 1997\\
Pio94 & Piotto et al. 1994 \\
Phe98 & Phelps et al. 1998 \\
Raw97 & Rawson et al. 1997 \\
Sah94 & Saha et al. 1994 \\
Sah95 & Saha et al. 1995 \\
Sa85a & Sandage \& Carlson 1985a \\ 
Sa85b & Sandage \& Carlson 1985b \\
Sa88a & Sandage 1988  \\
Sa88b & Sandage \& Carlson 1988 \\
Sh96a & Saha et al. 1996a \\
Sh96b & Saha et al. 1996b \\
Sh96c & Saha et al. 1996c \\
Sil96 & Silbermann et al. 1996 \\
Sil98 & Silbermann et al. 1998 \\
Tam68 & Tamman \& Sandage 1968 \\
Tan95 & Tanvir et al. 1995 \\
To95a & Tolstoy et al. 1995a \\
To95b & Tolstoy et al. 1995b \\
Tur98 & Turner et al. 1998 \\
Vis89 & Visvanathan 1989 \\
Wal88 & Walker 1988 \\
Wel86 & Welch et al. 1986 \\
\hline
\end{tabular}
\end{minipage}}
\end{center}
\end{table}

\begin{verbatim}
Table 6: Extract of the ASCII file for the Cepheid V1 of galaxy IC1613.
----------------------------------------
IC1613  V1             9
       1    .748     N
       2   21.36     mea       B   Fr88a
       2   20.79     mea       V   Fr88a
       2   20.36     mea       R   Fr88a
       2   20.14     mea       I   Fr88a
       2   20.50     max       B   Sa88a
       2   22.03     min       B   Sa88a
       2   21.27     ave       B   Sa88a
       2   21.39     mea       B   Sa88a
----------------------------------------
\end{verbatim}

\refer 
\aba
\rf{Mc Alary, C.W. et al.: 1983, ApJ. 273, 539}
\rf{Mc Alary, C.W., Madore, B.F., Davis,  L.E.: 1984, ApJ. 276, 487}
\rf{Alves, D.R., Cook, K.H.: 1995, AJ. 110, 192}
\rf{Baade, W., Swope, H.H.: 1963, AJ 68, 435}
\rf{Baade, W., Swope, H.H.: 1964, AJ 70, 212}
\rf{Carlson, G., Sandage, A.: 1990, ApJ. 352, 587}
\rf{Capaccioli, M., Piotto, G., Bresolin, F.: 1992, AJ. 103, 1151}
\rf{Christian, C.A., Schommer, R.A.: 1987, AJ. 93, 557}
\rf{Cook, K.H., Aaronson, M.: 1986, ApJ. Letters, 301, L45}
\rf{Di Nella-Courtois, H., Lanoix, P., Paturel, G.: 1999, MNRAS (in press)}
\rf{Ferrarese, L. et al.: 1996, ApJ. 464, 568}
\rf{Ferrarese, L. et al.: 1998, ApJ. 507, 655}
\rf{Freedman, W.L.: 1988, ApJ. 326, 691}
\rf{Freedman, W.L., Madore, B.F.: 1988, ApJ. Letters 332, L63}
\rf{Freedman, W.L., Madore, B.F.: 1990, ApJ. 365, 186}
\rf{Freedman, W.L., Wilson, C.D., Madore, B.F.: 1991, ApJ. 372, 455}
\rf{Freedman, W.L., Madore, B.F., Hawley, S.L., Horowitz, I.K., Mould, J., Navar
ette, M., Sallmen, S.: 1992, ApJ. 396, 80}
\rf{Freedman, W.L. et al.: 1994, ApJ. 427, 628}
\rf{Gallart, C., Aparicio, A., Vichez, J.M.: 1996, AJ. 112, 1928}
\rf{Gaposchkin, S.: 1962, AJ 67, 334}
\rf{Gibson B.K. et al.: astro-ph/981003}
\rf{Graham, J.A. et al.: 1997, ApJ. 477, 535}
\rf{Hoessel, J.G., Abbott, J., Saha, A., Mossman, A.E., Danielson, G.E.: 1990, A
J. 100, 1151}
\rf{Hoessel, J.G., Saha, A., Krist, J., Danielson, G.E.: 1994, AJ. 108, 645}
\rf{Hoessel, J.G., Saha, A., Danielson, G.E.: 1998, AJ. 115, 573}
\rf{Hughes, S.M.G. et al.: 1998, ApJ. 501, 32}
\rf{Kayser, S.E.: 1967, A.J. 72, 134}
\rf{Kelson, D.D. et al.: 1996, ApJ. 463, 26}
\rf{Lanoix, P.: 1998, AA 331, 421}
\rf{Lanoix, P., Paturel, G., Garnier, R.: 1999a, MNRAS, submitted}
\rf{Lanoix, P., Paturel, G., Garnier, R.: 1999b, ApJ. 516, in press}
\rf{Kinman, T.D., Mould, J.R., Wood, P.R.: 1987, AJ. 93, 833}
\rf{Madore, B.F.: 1985, Proc. IAU Colloquium  82, Cambridge
University press}
\rf{Madore, B.F., Mc Alary, C.W., Mc Laren, R.A., Welch, D.L., Neugebauer, G., Matthews, K.: 1985, ApJ. 294, 560} 
\rf{Madore, B.F., Welch, D.L., Mc Alary, C.W., Mc Laren, R.A.: 1987, ApJ. 320, 26}
\rf{Mc Alary, C.W., Madore, B.F.: 1984, ApJ. 282, 101}
\rf{Mould, J.R.: 1987, PASP. 99, 1127}
\rf{Musella, I., Piotto, G., Capaccioli, M.: 1997, AJ 114, 976}
\rf{Piotto, G., Capaccioli, M., Pellegrini, C.: 1994, Astron. Astrophys. 287, 371}
\rf{Phelps, R.L. et al.: 1998, ApJ. 500, 763}
\rf{Rawson, D.M. et al.: 1997, ApJ. 490, 517}
\rf{Saha, A., Labhardt, L., Schwengeler, H., Macchetto, F.D., Panagia, N., Sandage, A., Tammann, G.A.: 1994, ApJ. 425, 14}
\rf{Saha, A., Sandage, A., Labhardt, L., Schwengeler, H., Tammann, G.A., Panagia,
N., Macchetto, F.D.: 1995, ApJ. 438, 8}
\rf{Saha, A., Sandage, A., Labhardt, L., Tammann, G.A., Macchetto, F.D., Panagia,
N.: 1996a, ApJ. 466, 55}
\rf{Saha, A., Hoessel, J.G., Krist, J., Danielson, G.E.: 1996b, AJ. 111, 197}
\rf{Saha, A., Sandage, A., Labhardt, L., Tammann, G.A., Macchetto, F.D., 
Panagia, N.: 1996c, ApJ. SS. 107, 693}
\rf{Sandage, A., Carlson, G.: 1985a, AJ. 90, 1464}
\rf{Sandage, A., Carlson, G.: 1985b, AJ. 90, 1019}
\rf{Sandage, A.: 1988, PASP. 100, 935}
\rf{Sandage, A., Carlson, G.: 1988, AJ. 96, 1599}
\rf{Silbermann, N.A. et al.: 1996, ApJ. 470, 1}
\rf{Silbermann, N.A. et al.: Astro-ph 9806017}
\rf{Tamman, G.A., Sandage, A.: 1968, ApJ. 151, 825}
\rf{Tanvir, N.R., Shanks, T., Ferguson, H.C., Robinson, D.R.T.: 1995, Nature 337, 27} 
\rf{Tolstoy, E., Saha, A., Hoessel, J.G., McQuade, K.: 1995a, AJ. 110, 1604}
\rf{Tolstoy, E., Saha, A., Hoessel, J.G., Danielson, G.E.: 1995b, AJ. 109, 579}
\rf{Turner A. et al.: 1998, ApJ. 505, 207}
\rf{Visvanathan, N.: 1989, ApJ. 346, 629}
\rf{Walker, A.R.: 1998, PASP 100, 949}
\rf{Welch, D.L., McAlary, C.W., Mclaren, R.A., Madore, B.F.: 1986, ApJ. 305, 583
}
\rf{}
\abe

\addresses
\rf{P. Lanoix, CRAL Observatoire de Lyon, F69230 Saint-Genis Laval, FRANCE,
Lanoix@obs.univ-lyon1.fr}
\rf{R. Garnier, CRAL Observatoire de Lyon, F69230 Saint-Genis Laval, FRANCE,
Garnier@obs.univ-lyon1.fr}

\end{document}